\documentclass[aps,prl,twocolumn,superscriptaddress,nofootinbib]{revtex4-1}

\usepackage[latin1]{inputenc}
\usepackage{graphicx}
\usepackage{xcolor}

\usepackage{amsmath}

\definecolor{darkblue}{rgb}{0,0,0.6}
\definecolor{darkred}{rgb}{0.6,0,0}
\usepackage[colorlinks=true,urlcolor=darkblue,citecolor=darkblue,linkcolor=darkred,hyperfootnotes=false]{hyperref}

\usepackage{ulem}

\begin{document}

%\title{Facetting and wrinkling in twisted ribbons}
\title{From cylindrical to stretching ridges and wrinkles in twisted ribbons}

\author{Huy Pham Dinh}
\affiliation{Laboratoire Interfaces $\&$ Fluides Complexes, Universit\'e de Mons, 20 Place du Parc, B-7000 Mons, Belgium}
\author{Vincent D\'emery}
%\email{vincent.demery@espci.fr}
\affiliation{UMR Gulliver 7083, CNRS and ESPCI Paris, 10 rue Vauquelin, 75005 Paris, France}
\author{Benny Davidovitch}
\affiliation{Department of Physics, University of Massachussetts, Ahmerst, MA 01003.}
\author{Fabian Brau}\thanks{Present address: Universit\'e libre de Bruxelles (ULB), Nonlinear Physical Chemistry Unit, CP231, 1050 Brussels, Belgium}
\affiliation{Laboratoire Interfaces $\&$ Fluides Complexes, Universit\'e de Mons, 20 Place du Parc, B-7000 Mons, Belgium}
\author{Pascal Damman}
%\email{pascal.damman@umons.ac.be}
\affiliation{Laboratoire Interfaces $\&$ Fluides Complexes, Universit\'e de Mons, 20 Place du Parc, B-7000 Mons, Belgium}

\date{\today}

\begin{abstract}
Twisted ribbons subjected to a tension exhibit a remarkably rich morphology, from  smooth and wrinkled helicoids, to cylindrical or faceted patterns. These shapes are intimately related to the instability of the natural, helicoidal symmetry of the system, which generates both longitudinal and transverse stresses, thereby leading to buckling of the ribbon. In this paper, we focus on the tessellation patterns made of triangular facets. Our experimental observations are described within an ``asymptotic isometry'' approach that brings together geometry and elasticity. The geometry consists of parametrized families of surfaces, isometric to the undeformed ribbon in the singular limit of vanishing thickness and tensile load. The energy, whose minimization selects the favored structure among those families, is governed by the tensile work and bending cost of the pattern. This framework describes the coexistence lines in a morphological phase diagram, and determines the domain of existence of faceted structures.
\end{abstract}

\maketitle

Sheets subjected to external forces store the exerted work in local or global elastic deformations that underlie wrinkled and crumpled states. For tensile loads, the exerted work is typically stored as pure stretching energy. When the forces are compressive, the strain is negligible and the exerted work is instead stored as bending energy. For instance, a sheet resting on a soft substrate and compressed uniaxially deforms isometrically into wrinkles or folds~\cite{pocivavsek_08,brau11}. However, when the compressive forces result from geometrical constraints, the final shape may involve a complex combination of bending and stretching energies. For instance, a complex 3D shape made by confining a thin sheet, such as a crumpled paper ball, is often described as an assembly of flat polygonal facets delimitated by ridges where stretching and bending predominate~\cite{witten_07}. Such a faceted morphology is an efficient minimizer of stretching since it is isometric to the undeformed sheet ({\textit{i.e.}} it is strainless) everywhere except at those narrow ridges, reaching a perfect isometry only at the singular limit of vanishing thickness.

Two types of ridges have been reported: {\it i)} isometric (cylindrical) ridges, which involve only bending, and {\it ii)} stretching ridges, in which the bending and stretching energies are comparable, leading to a width $w_{\mathrm{r}} \sim L^{2/3}t^{1/3}$, where $L \gg w_{\mathrm{r}}$ is the length of the ridge and $t\ll w_{\mathrm{r}}$ is the thickness of the sheet \cite{witten_07,audoly_10}. Transitions from isometric to stretching ridges were reported for very simple geometries. Witten showed that a single stretching ridge becomes isometric when both ends are truncated~\cite{witten_09}, and Fuentealba \textit{et al.}\ have studied a tearing flap subjected to a pulling force and found a transition from isometric to stretching ridge when the pulling force exceeds a threshold $F_{\mathrm{c}}\sim B/(t^{2/3}L^{1/3})$ (where $B\sim E t^3$ is the bending modulus and $E$ the Young's modulus of the sheet)~\cite{fuentealba_15}. Both studies suggest that the curvature imposed at the end of the ridge completely determines its shape.

\begin{figure}
\begin{center}
\includegraphics[width=.75\linewidth]{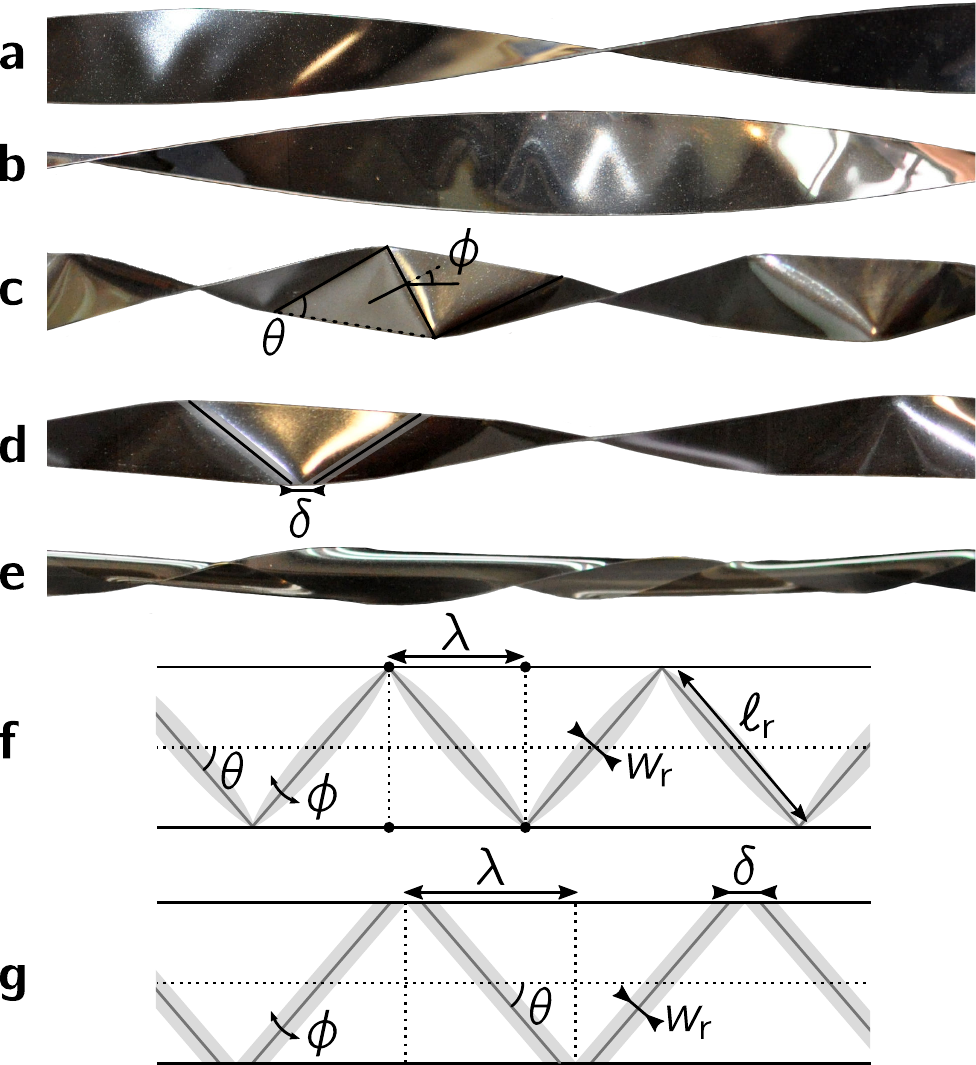}
\end{center}
\caption{{\bf a-e}: Pictures of the different morphologies: {\bf a.} helicoid, {\bf b.}  wrinkled helicoid, {\bf c.} stretching ridges, {\bf d.} isometric ridges, {\bf e.} cylinder. {\bf f.} Parametrization of the stretching ridges reported on {\bf c}, the gray areas denote the ridges. {\bf g.} Parametrization of the isometric ridges reported on {\bf d}.}
\label{fig:photos_schemes}
\end{figure}

In this letter, we investigate transitions between isometric and stretching ridges in the complex morphology of a twisted ribbon, and characterize their impact on their mechanics. The classical set-up of a twisted ribbon whose short edges are held apart by a small tensile force was first studied by Green~\cite{green_36,green_37}. 
It was recently revisited by Chopin and Kudrolli~\cite{chopin_13}, who proposed to organize the observed morphologies in a tension-twist phase diagram. The helicoid (Fig.~\ref{fig:photos_schemes}a), characterized by a stretching of the edges as well as the midline, appears usually for moderate twist angles and large tensions. Increasing the twist at moderate tension, the helicoid undergoes a buckling instability, whereby longitudinal wrinkles also described as a zig-zag fold form around its midline~\cite{coman_08} (see Fig.~\ref{fig:photos_schemes}b).%\footnote{\textcolor{blue}{This is the second shape we describe in the introduction whereas this is the last shape discussed in the paper. May be we should follow in the introduction the experimental trajectory showed by gray arrows in Fig.2 since we follow this trajectory in the rest of the text. This is more logical (FIR $\to$ FSR $\to$ WH $\to$ cylinder). I have already switched the position of the isometric and stretched ridge and adapted the references in the text. But before considering this comment, please read the next footnote.}}. 
Further increasing the twist leads to a faceted morphology, also called ``creased helicoid''~\cite{chopin_13} or ``ribbon crystal''~\cite{bohr_13} (Figs.~\ref{fig:photos_schemes}c and ~\ref{fig:photos_schemes}d). Additionally, the helicoid buckles in the transverse direction upon increasing the twist at sufficiently large tension~\cite{chopin_15}, and reaches a cylindrical shape (Fig.~\ref{fig:photos_schemes}e). We must clarify here that in the above sentences and throughout this letter we refer to tensile loads as  ``large'', ``moderate'', or ``small'' according to their ratio with certain powers of the thickness $t$, but even the ``large tension'' corresponds to characteristic strains of $10^{-2}$ or less, deeply in the Hookean regime of material response. In this sense, the morphological transitions reported in~\cite{green_36,green_37,chopin_13}, as well as in this work, are universal and not material-dependent.   

The faceted morphology has been described using a purely isometric shape, either by solving effective equations for the midline of the ribbon~\cite{korte_11} or by assuming that it consists of flat triangular facets separated by isometric ridges~\cite{bohr_13}. However, facets are observed over a whole region of the twist-tension phase diagram, where the tension is small but nonzero~\cite{chopin_13}, suggesting that this morphology can accommodate a finite amount of stretching. Furthermore, upon increasing the tension at moderate twist, facets turn into longitudinal wrinkles, which are clearly stretched~\cite{chopin_13} (see Fig.~\ref{fig:photos_schemes}b). Motivated by these observations, we focus here on the faceted morphology: we determine experimentally its domain of existence and propose a theoretical framework to explain how facets separated by isometric ridges (FIR) that are observed at small tension, turn into facets separated by stretching ridges (FSR), and then to longitudinal wrinkles (WH) at large tension.

\begin{figure}[!t]
\includegraphics[width=0.45\textwidth]{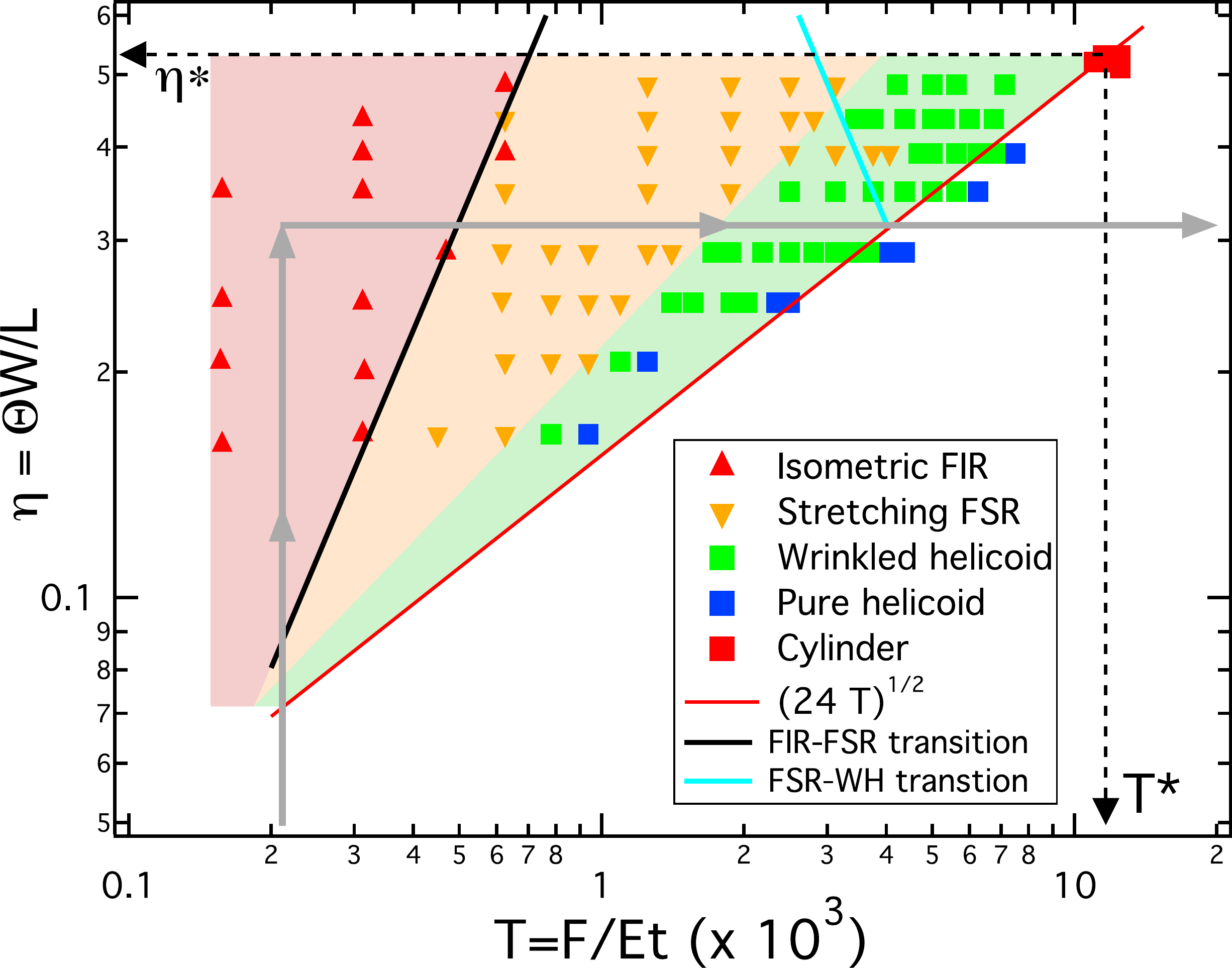}
\caption{Phase diagram of a PET ribbon with $t \simeq 0.012$ and $L \simeq 18$. The solid lines correspond to the co-existence lines (see text). The gray lines with arrows represent the experimental trajectory. $T^*$ and $\eta^*$ corresponds to the helicoid-cylinder transition at fixed length~\cite{chopin_13}.}
\label{phase_diag}
\end{figure}

We use ribbons of length $L$, width $W$, and thickness $t$ under external tension $T$ and clamped at their short edges, which are twisted relative to each other by a prescribed angle $\Theta$. Our ribbons are composed of polyethylene terephthalate PET (Young modulus $E \simeq 3\,\textrm{GPa}$) or softer polydimethyl siloxane PDMS ($E \simeq 1\,\textrm{MPa}$). For simplicity of the analysis, we assume the Poisson ratio $\nu=0$ (a $\nu \neq 0$ analysis has no qualitative implications). We use $W$ as a unit of length, and the stretching modulus $Y=tE$ as a unit of in-plane stress (\textit{i.e.}, $W\!=\!Y\!=\!1$), and introduce the twist per unit ribbon length $\eta\!=\!\Theta W/L$, and the energy per unit length $U$. The different observed shapes are shown on Fig.~\ref{fig:photos_schemes} and are organized in a phase diagram (Fig.~\ref{phase_diag}), which focuses on the longitudinally buckled morphologies and complements the one reported in~\cite{chopin_13}.

\begin{figure}[!b]
\includegraphics[width=0.45\textwidth]{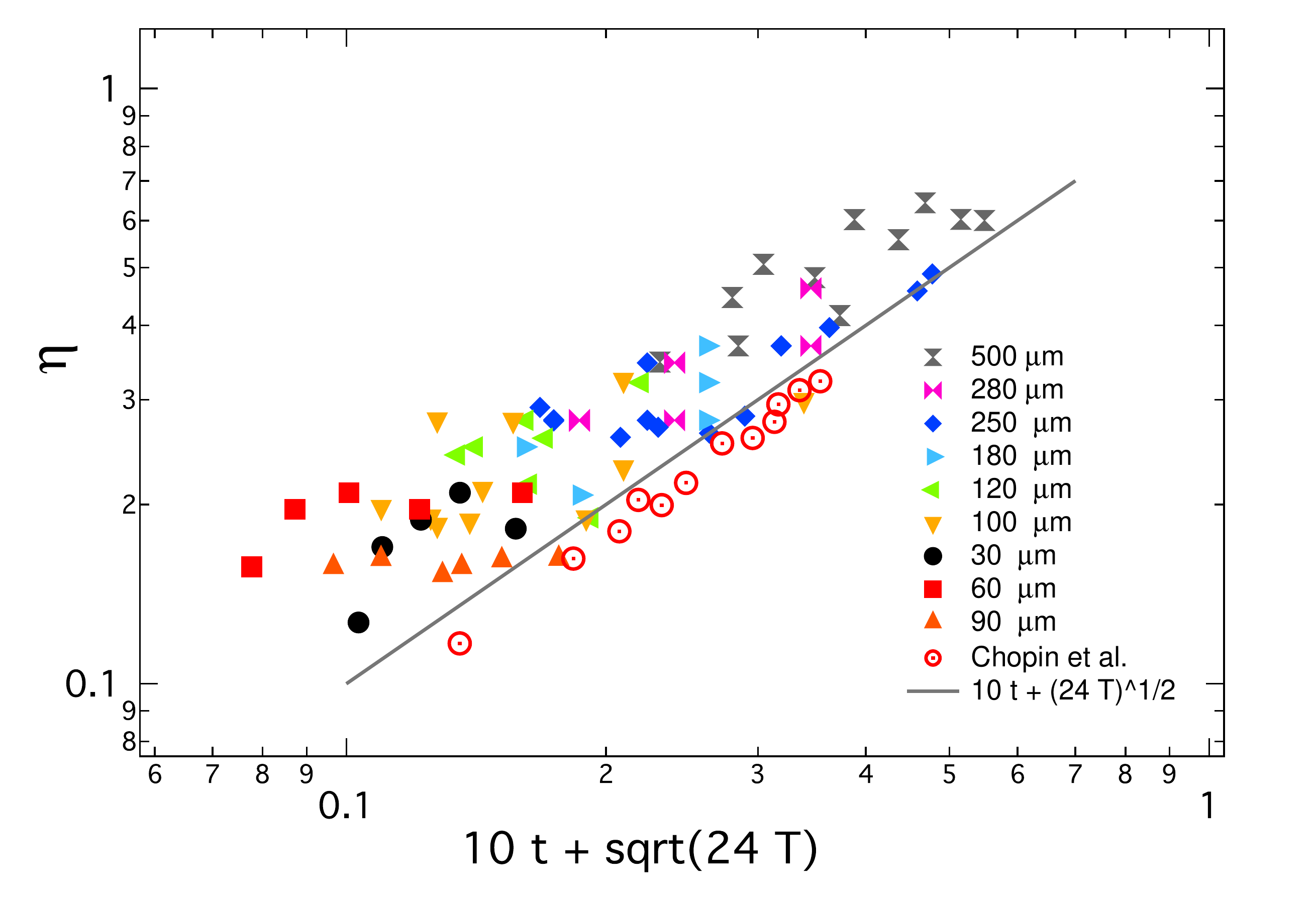}
\caption{Critical twist angle for the helicoid-facets transition, $\eta_{\ell} $ as a function of $\sqrt{24 T} + 10 t$ for different ribbons of PET (thicknesses are as indicated). 
}
\label{fig:lambda_T_eta}
\end{figure}

Our experimental ``trajectory'' is depicted by the gray lines in Fig.~\ref{phase_diag}. To avoid hysteresis, the tension at each segment is either constant or increases. We start at a very small tension and zero twist, and progressively increase the twist. Once the chosen twist angle is attained, the tension is increased progressively. First, the ribbon takes a helicoidal shape (Fig.~\ref{fig:photos_schemes}a). As $\eta$ is increased further, such that $\eta>\sqrt{24 T}$, the centerline of the helicoid is under compression. A linear stability analysis shows that buckling occurs when $\eta_{\ell} \simeq \sqrt{24 T} +10t$, forming first wrinkled helicoids~\cite{coman_08,green_36,green_37,chopin_13,chopin_15}. The prediction for the critical twist is compared to our experimental results in Fig.~\ref{fig:lambda_T_eta}, and shown in the phase diagram Fig.~\ref{phase_diag} (red line).

{\bf Facets separated by isometric ridges (FIR).}
Increasing the twist beyond the buckling threshold, still at very small tension, we observe a shape resembling facets separated by rounded cylindrical ridges (see Fig.~\ref{fig:photos_schemes}d). We follow Ref.~\cite{bohr_13} and model it with flat triangular facets separated by cylindrical ridges. Such a shape is parametrized by the two angles $\theta$ and $\phi$, formed, respectively, between the ridges and the ribbon's midline, and between adjacent facets (Fig.~\ref{fig:photos_schemes}c and~\ref{fig:photos_schemes}g), and by the radius of curvature of the ridges, $R_{\mathrm{c}}$. In contrast to Ref.~\cite{bohr_13}, we find $\theta$ and $R_{\mathrm{c}}$ by energy minimization.

We evaluate the energy using the general framework of \textit{asymptotic isometries}~\cite{chopin_15}, which is valid for physically admissible states in the doubly-asymptotic limit of vanishing tension $T$ and thickness $t$. The energy of such states can be approximated by the sum of a tensile work and an elastic energy $U^\mathrm{el}$, both of which vanish in the limit $t,T \to 0$: 
\begin{equation}\label{eq:ai}
U=U^\mathrm{el}+\chi T,
\end{equation}
where the contraction is $\chi=1-L_{\mathrm{ee}}/L$ ($L_{\mathrm{ee}}$ being the end-to-end distance). This formalism allows to compute both $U^\mathrm{el}$ and $\chi$ from geometrical and mechanical considerations, yielding expressions that have no explicit dependence on $T$ and can be formally evaluated at $T=0$. We can thus make a tension-independent construction by using parameters with geometrical meaning ($\theta, \phi, R_{\mathrm{c}}$), whose actual dependence on $T$ is found when minimizing the whole energy, Eq.~(\ref{eq:ai}). Notice that the (unwrinkled) helicoid, for instance, is not an asymptotic isometry of the ribbon, since its elastic energy is proportional to $\eta^4$ and does not vanish as $t,T \to 0$~\cite{chopin_15}. 
More generally, in the asymptotic limit $t \to 0$, the AI framework is valid only if ratio $eta^2/T$, between the  twisted-induced stress $\eta^2$, and the exerted tension $T$, is sufficiently large.
We will use this framework not only for the FIR but also for the other shapes obtained upon increasing the tension. A similar approach has been used in other studies of sheets (or shells) on which a Gaussian curvature is imposed in the presence of small tension~\cite{vella_15,vella_15b,paulsen_15}. 

The twist $\eta(\theta,\phi,R_{\mathrm{c}})$ and contraction $\chi(\theta,\phi,R_{\mathrm{c}})$ of a FIR can be obtained from geometrical arguments (see Ref.~\cite{bohr_13} and SM~\cite{SM}). Upon expansion for small $\eta$ and $R_{\mathrm{c}}$, we obtain:  
\begin{equation}\label{eq:chi_facets}
\chi_{\mathrm{FIR}} = \frac{\eta^2}{8} + \frac{\eta^3 R_{\mathrm{c}}}{\sin(2\theta)}+ \left[\frac{1}{48\sin(\theta)^2}-\frac{5}{384} \right]\eta^4 + \mathcal{O}(\eta^5 R_{\mathrm{c}}).
\end{equation}

The bending energy of a single ridge is given by \mbox{$u_{\mathrm{r}}\sim t^2 \eta/[R_{\mathrm{c}}\sin(\theta)^2]$} (using that $\phi\simeq \eta/\sin(\theta)+\mathcal{O}(\eta^3)$ \cite{SM}). Assuming a small width of the ridge compared to the wavelength $w_{\mathrm{r}}\ll\lambda$, there are $N/L \simeq 1/\lambda=\tan(\theta)$ ridges per unit ribbon length, and the elastic energy per unit ribbon length becomes
\begin{equation}
U^\mathrm{el}_{\mathrm{FIR}}\sim \frac{t^2\eta}{R_{\mathrm{c}}\sin(2\theta)}.
\end{equation}

\begin{figure}
\includegraphics[width=0.45\textwidth]{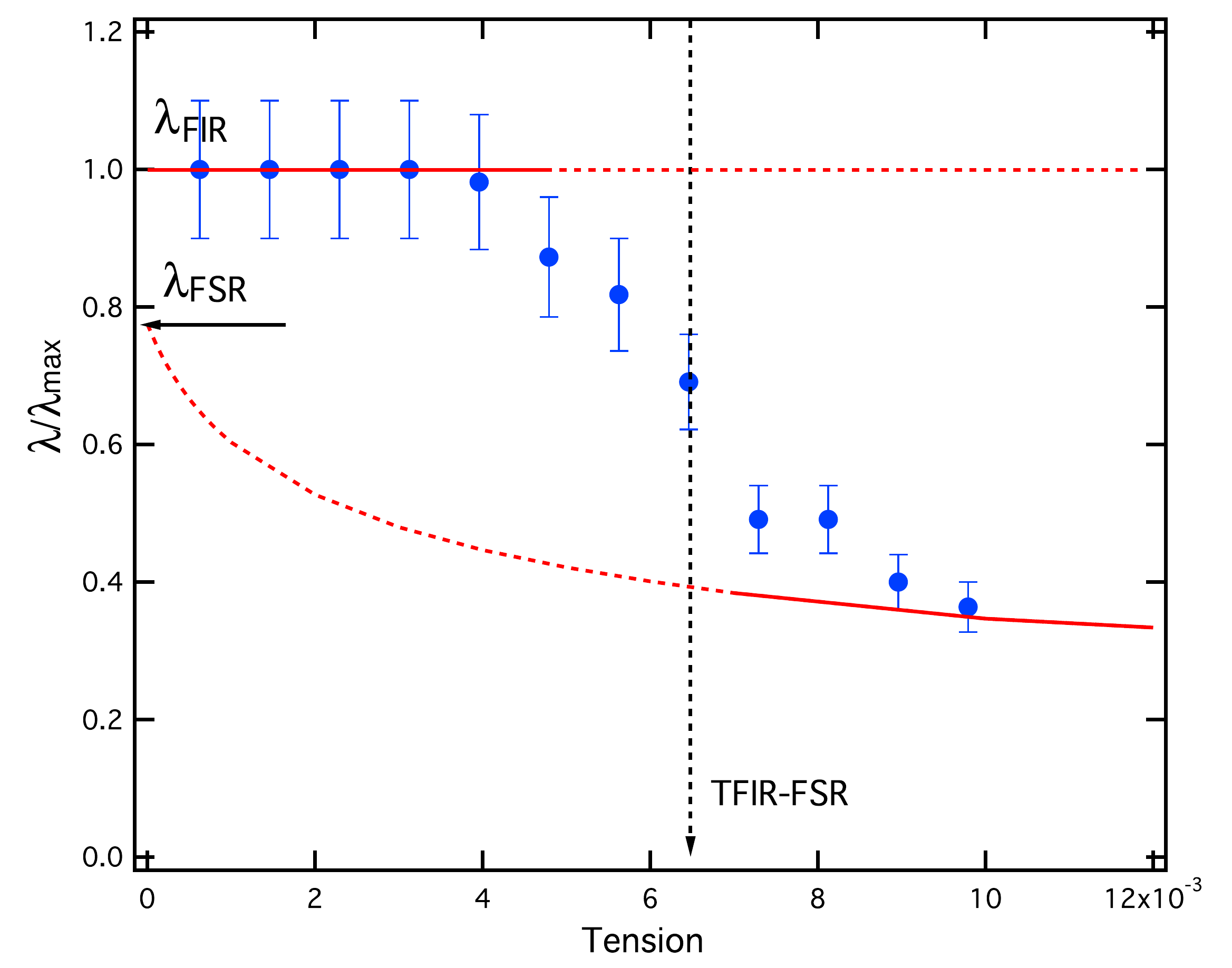}
\caption{Evolution of the measured size of the facets $\lambda/\lambda_{\text{max}}$ (filled symbols) for different PET ribbons ($\lambda_{\text{max}}\simeq 1.3$, $t=0.012, \eta = 0.37$). The red lines corresponds to the theoretical $\lambda(T)$ curves for FIR and FSR.}
\label{fig:lambda_T}
\end{figure}

Minimizing the global energy $U^\mathrm{el}_{\mathrm{FIR}}+\chi_{\mathrm{FIR}} T$ (keeping terms up to the order $\eta^3$ in the contraction) yields: 
\begin{gather}\label{eq:FIR1}
\theta  = \pi/4, \quad \lambda=1, \quad  R_{\mathrm{c}} \sim t/(\eta\sqrt{T}), \\
U_{\mathrm{FIR}}\sim t\eta^2\sqrt{T}+ \eta^2 T  + \mathcal{O}(\eta^4). \label{eq:FIR2} 
\end{gather}
Notably, the independence of the wavelength $\lambda$ on tension indicates the robust, geometrical nature of the FIR shape, despite the nontrivial dependence of its energy on the tensile load $T$. In hindsight, this robustness explains the validity of the purely geometric approach of Refs.~\cite{korte_11,bohr_13} for describing the general structure of the FIR. In Fig.~\ref{fig:lambda_T}, we test the prediction $\lambda=1$ by varying the tension $T$ at constant $\eta$. Only the low-$T$ regime, where $\lambda$ is constant, exhibits the FIR. The independence of $\lambda$ on $T$ is in agreement with the theory, despite a slight discrepancy between the observed and predicted values of the wavelength that may be attributed to the finite size of the ribbon ($L\simeq 15$). We also note that a similar scaling of $R_{\mathrm{c}}$ with $t$ and $T$ was obtained in Ref.~\cite{fuentealba_15} for a completely different system, namely the width of a pulling flap in an isometric configuration. Finally, inspection of Eq.~(\ref{eq:FIR2}) shows that the tensile work and elastic energy are balanced only for $T \sim t^2$. Therefore, for FIR at very small tension ($T \ll t^2$), the work done by the torque upon twisting the ribbon is stored efficiently as bending energy in the ridges, whereas observation of FIR for larger tension ($T \gg t^2$) implies that the twister transmits its work to the puller and the ribbon becomes a ``bad capacitor'' of energy.

%%%%%%%%%%%%%%%%%%%%%%%%%%%%%%
{\bf Facets separated by stretching ridges (FSR).}
Upon increasing the tension, the radius of curvature of the ridges decreases, until the ridges pinch along the ribbon's long edges (Fig.~\ref{fig:photos_schemes}c). From visual inspection and the study of pulling flaps~\cite{fuentealba_15}, we hypothesize that the ribbon's shape consists of facets separated by stretching ridges (FSR).

For the FSR, the contraction $\chi_{\mathrm{FSR}}$ can be directly evaluated by considering $R_{\mathrm{c}} \ll \eta$~\cite{note_Rc}, retaining only the $\eta^2$ and $\eta^4$ terms of Eq.~(\ref{eq:chi_facets}). The elastic energy of a single stretching ridge is given by $u_{\mathrm{r}}\sim t^{5/3}\ell_{\mathrm{r}}^{1/3}\phi^{7/3}$, where $\ell_{\mathrm{r}}=1/\sin(\theta)$ is the length of a ridge~\cite{witten_95,lobkovsky_96}. Using the above geometric relationships between  $\phi, \lambda, \theta$ and $\eta$, we deduce that the elastic energy per unit ribbon length is: 
\begin{equation}\label{eq:mf_uel}
U_{\mathrm{FSR}}^\mathrm{el}\sim \frac{t^{5/3}\eta^{7/3}}{\sin(\theta)^{5/3}\cos(\theta)}.
\end{equation}
Again, the total energy $U^\mathrm{el}_{\mathrm{FSR}}+\chi_{\mathrm{FSR}} T$ should be minimized. The angle $\theta$ is the solution of 
%We find that, contrary to the FIR, the wavelength of FSR does depend on the tensile load.
\mbox{$8(3\tan(\theta)^2 - 5) \sin(\theta)^{1/3} / \cos \theta=(\eta/t)^{5/3}T$}. Interestingly, a physical solution only exists for $\theta \ge \theta_c = \arctan(\sqrt{5/3})$ which determines the size of facets at vanishing tension, $\lambda = \sqrt{3/5}\simeq 0.78$. 
For small tension, the wavelength slightly decreases with tension as $\lambda \simeq 0.78 - 6.4 \, 10^{-3} \, (\eta/t)^{5/3}T$.
%For vanishing tension, $\theta = \arctan (\sqrt{5/3}) \simeq 52^\circ$, or $\lambda = \sqrt{3/5} \simeq 0.78$.
The energy of the FSR ribbon at small tension becomes,
\begin{gather}
U_{\mathrm{FSR}} \sim t^{5/3} \eta^{7/3} + \eta^2 T  + \mathcal{O}(\eta^4). \label{eq:FSR2} 
\end{gather}
For isometric ridges (FIR), increasing the tension decreases the radius of curvature, $R_{\mathrm{c}}$, of the ridges, Eq.~(\ref{eq:FIR1}), thus increasing the elastic energy of the ribbon. At some critical value of the tension, it becomes energetically favourable to switch to stretching ridges (FSR) which enable saving some bending energy. Comparing the total energy of both faceted shapes, Eqs. (\ref{eq:FIR2}) and (\ref{eq:FSR2}), we find that the FSR appears for tensions above 
\begin{equation}\label{eq:t_fir_fsr}
T_{\mathrm{FIR-FSR}} \sim t^{4/3}\eta^{2/3}.
\end{equation}
This prediction is shown in the phase diagram, Fig.~\ref{phase_diag}, in good agreement with our experimental observations. This scaling can also be obtained through a direct comparison of the widths of isometric, $\eta R_{\mathrm{c}}\sim t/\sqrt{T}$, and stretching, $(t/\eta)^{1/3}$ ridges. We note that the predicted dependence of the tension in $t$ is similar to the transition force found in~\cite{fuentealba_15} between isometric and stretching ridges for pulling flaps. Figure~\ref{fig:lambda_T} provides a significant support for the theoretical prediction of a sharp transition of the wavelength $\lambda$, from a tension-independent plateau in a low-$T$ regime (FIR), to a tension-dependent branch (FSR). Another recent work~\cite{chopin_16b} also found a sharp transition of the facet's size upon increasing tension. 

In the FSR regime, the evolution of the wavelength as a function of the tension can be obtained numerically up to an unknown numerical factor multiplying $T$ (see Fig.~\ref{fig:lambda_T}). To determine the asymptotic behavior of the wavelength $\lambda$ at large tension, we expand the two terms in the energy in $\lambda\simeq (\pi/2)-\theta$:
\begin{equation}\label{eq:ufsr_large_tension}
U_{\mathrm{FSR}}^\mathrm{el}+\chi_{\mathrm{FSR}}T\sim \frac{t^{5/3}\eta^{7/3}}{\lambda}+T\eta^2 \left[1+\eta^2\left(1+\lambda^2\right) \right],
\end{equation}
whose minimization leads to $\lambda\sim (t/\eta)^{5/9}T^{-1/3}$. This relation is consistent with the general trend for the wavelength. Inserting the wavelength expression in Eq.~(\ref{eq:ufsr_large_tension}) yields the total energy of the FSR ribbon for large tensions, \textit{i.e.} when the angle $\theta$ is close to $\pi/2$,
\begin{equation}\label{eq:ufsr_large_tension2}
U_{\mathrm{FSR}}^{T} \sim t^{10/9} \eta^{26/9} T^{1/3} + T \eta^2.
\end{equation}
The FSR description assumes that the width of the ridges, which is given by $w_{\mathrm{r}}\sim (t/\eta)^{1/3}$, remains small compared to the size of the facets. This assumption holds as long as $T<(t/\eta)^{2/3}$. 

%%%%%%%%%%%%%%%%%%%%%%%%%%
{\bf Wrinkled helicoid (WH).} 
Turning now to higher tension values, it is natural to ask how the FSR state (Fig.~\ref{fig:photos_schemes}c) transforms into the wrinkled helicoid (Fig.~\ref{fig:photos_schemes}b), observed for tensions slightly smaller than the fixed ribbon length limit, $T=\eta^2/24$ (see Fig.~\ref{phase_diag})\cite{contraction}. 
For $T < \eta^2/24 $, the stress field within a twisted ribbon can be divided into three parts. The central part of the helicoid $|r|<r_{\mathrm{wr}}$ is under compression, while the two outer parts ($r_{\mathrm{wr}}< |r| < 1/2$) remain stretched. A basic description of the wrinkled helicoid state has been developed in \cite{chopin_15}, using a far-from-threshold theory, whereby the inner zone around the helicoid's midline is decorated with longitudinal wrinkles that fully relax compression while the two outer strips are stretched. The width $2r_{\mathrm{wr}}$ of the wrinkled zone is determined by the ratio $T/\eta^2$, vanishing for $T/\eta^2 \to 1/24$ and close to $1$ at asymptotic isometry, where $ T/\eta^2 \to 0$. In this limit, the non-developable wrinkled helicoid provides a remarkable example of an isometry that does not correspond to a piecewise-developable shape (in contrast to the faceted shapes that we discussed above). While a full characterization of the wrinkled helicoid is beyond the scope of the current work, we use below a scaling analysis to explore the possibility of a transition from FSR to a wrinkled helicoid at the asymptotic isometry limit $t, T \to 0$. Note that this is obviously a crude approximation of the shape observed in Fig.~\ref{fig:photos_schemes}b, where the wrinkles do not reach the edges. %The presence of stretched outer part in the ribbon is not compatible with the AI framework.

Assuming that the ribbon does approach a fully wrinkled helicoid shape at $T \ll \eta^2$, the contraction $\chi_{\mathrm{WH}}$ can be computed by assuming that the edges, $|r|=r_{\mathrm{wr}}\approx 1/2$, are neither wrinkled nor stretched. A simple calculation yields
\begin{equation}\label{eq:contraction_wr}
\chi_{\mathrm{WH}}=1-\sqrt{1-\frac{\eta^2}{4}}=\frac{\eta^2}{8}+\frac{\eta^4}{128}+\mathcal{O}(\eta^6).
\end{equation}
Note that this expression corresponds to the contraction of the facets given by Eq.~(\ref{eq:chi_facets}), in the limit $R_{\mathrm{c}}\to 0$, $\theta\to\pi/2$. 
In order to evaluate the elastic energy, we must determine the amplitude $A$ and wavelength $\lambda$ of the wrinkles. Since the wrinkles completely relax the compression, we obtain the ``slaving condition'': $A/\lambda\sim\eta$~\cite{chopin_15}. The elastic energy is governed by stretching in the transverse direction, $U_{\mathrm{str}}\sim A^4$, and bending in the longitudinal direction, $U_{\mathrm{bend}}\sim t^2 A^2/\lambda^4$. Minimizing the total energy with respect to $\lambda$ yields: 
\begin{equation}
\lambda  \sim (t/\eta)^{1/3}, \quad U_{\mathrm{WH}} \sim t^{4/3}\eta^{8/3} + T \eta^2. \label{eq:WH}
\end{equation}
Unsurprisingly, the wavelength scales with the thickness similarly to the width of the stretching ridge~\cite{witten_95}, reflecting the same type of energy balance used in both cases. Comparing the energy estimates for both morphologies, Eqs.~(\ref{eq:ufsr_large_tension2}) and (\ref{eq:WH}), we see that the WH is energetically favorable for tensions above
\begin{equation}\label{eq:t_fsr_wh}
T_{\mathrm{FSR-WH}}\sim(t/\eta)^{2/3}.
\end{equation} 
Remarkably, this occurs when the size of the facets becomes comparable to the width of a ridge in the FSR. Note that this value is much larger than the tension where FSR appears, see Eq.~(\ref{eq:t_fir_fsr}), which guarantees a large domain of existence for the FSR. 
However, the predicted value of the transition $T_{\mathrm{FSR-WH}}$ is larger than the critical tension $T^*\sim t$ (see Fig.~\ref{phase_diag}) where transverse buckling instability occurs~\cite{chopin_15}, meaning that the WH cannot exist in the asymptotic isometry limit.
%
%This is not incompatible with our observations, because our experimental data for $t =0.012$ shows a transition at $T/\eta^2 \approx 1/40$ (Fig.~\ref{phase_diag}), rather close to the onset of wrinkles (at $T/\eta^2 <1/24$) and thus far from the asymptotic isometry regime\footnote{\textcolor{red}{I don't understand this sentence.}}.
The experimental data shows a transition at $T/\eta^2 \simeq 1/40$ (Fig.~\ref{phase_diag}), rather close to the onset of longitudinal buckling, $T/\eta^2 <1/24$. We are thus far from the asymptotic isometry regime ($T/\eta^2 \to 0$).

In conclusions, we employed here the framework of asymptotically isometric shapes, together with an analysis of the energy of elastic ridges, to classify the various types of longitudinally buckled morphologies attained by a twisted ribbon at very small thickness and tensile load: facets separated by isometric or stretching ridges (FIR, FSR), and wrinkled helicoid. 
We hope that the rich plethora of distinct patterns and transitions, will lead to further studies of this system as a model for the spontaneous emergence of morphological complexity in elastic sheets under geometric constraints.

\begin{acknowledgments}
The authors thank J.~Chopin and M.~Adda-Bedia for fruitful discussions, and
J. Bohr and S. Markvorsen for helpful clarifications about their ribbon crystal computations.
S. Cuvelier is acknowledged for technical assistance.
This work was partially supported by a grant from the Belgian CUD program, and by NSF CARRER award DMR 11-51780 (BD).
\end{acknowledgments}

%\bibliographystyle{unsrt}
%\bibliography{Facets_twisted_ribbons_PRL}{}

\newpage

\begin{widetext}

%%\documentclass[aps,twocolumn,nofootinbib]{revtex4}
%%\documentclass[aps,preprint]{revtex4}
%\documentclass[aps,prl,onecolumn,superscriptaddress]{revtex4-1}

%\usepackage[latin1]{inputenc}
%%\usepackage[T1]{fontenc}
%\usepackage{graphicx}
%\usepackage{xcolor}
%\usepackage{amsmath}

%%\textwidth 183mm
%%\textheight 247mm
%%\oddsidemargin -10mm
%%\headsep 10mm

%%\usepackage{xr}
%%\externaldocument{Facets_twisted_ribbons_PRL_V6}

%\definecolor{darkblue}{rgb}{0,0,0.6}
%\definecolor{darkred}{rgb}{0.6,0,0}
%\usepackage[colorlinks=true,urlcolor=darkblue,citecolor=darkblue,linkcolor=darkred,hyperfootnotes=false]{hyperref}

%\newcommand{\fig}{{Figure~}}
%\newcommand{\iden}{{\bf 1}}
\newcommand{\ind}[1]{_{\mathrm{#1}}}

\renewcommand{\aa}{\boldsymbol{a}}
\newcommand{\ee}{\boldsymbol{e}}
\newcommand{\pp}{\boldsymbol{p}}
\newcommand{\vv}{\boldsymbol{v}}
%\newcommand{\xx}{\boldsymbol{x}}
%\newcommand{\XX}{\boldsymbol{X}}
%\newcommand{\zz}{\boldsymbol{z}}

%%\DeclareMathOperator{\tr}{Tr}
%%\DeclareMathOperator{\sinc}{sinc}
%%\renewcommand{\div}{\mathrm{div}\,}
%\newcommand{\dd}{\mathrm{d}}

%\newcommand{\red}[1]{\textcolor{red}{#1}}
%\newcommand{\afaire}[1]{\red{\bf #1}}

%\newcommand{\etal}{\emph{et al.}}

%\begin{document}

%\title{From cylindrical to stretching ridges and wrinkles in twisted ribbons\\ Supplemental Information}

\begin{center}
\large \bf From cylindrical to stretching ridges and wrinkles in twisted ribbons\\ Supplemental Material
\end{center}

\section{Facets separated by stretching ridges (FSR)}

\subsection{Twist and contraction}\label{sm:facettes}

We compute the twist and contraction of the facetted morphology as a function of the two angles $\theta$ and $\phi$ that parametrize it.
A facetted ribbon of unit width is parametrized by two angles (see Fig.~\ref{fig:scheme}):
\begin{itemize}
\item the angle $\theta$ between the ridges and the centerline,
\item the angle $\phi$ between two adjacent facets.
\end{itemize}
We compute the geometric properties of the ribbon:
\begin{itemize}
\item the twist per unit length $\eta$,
\item the contraction $\chi$ that is the relative difference between the rest length and the end-to-end distance of the twisted ribbon ($\chi>0$ when the ribbon is contracted).
\end{itemize}

\subsubsection{General expressions}

We follow the parametrization given in Ref.~\cite{bohr_13}, in the particular case of zero curvature radius of the folds (the curvature radius at both ends of a stretching ridge is zero).
The unit cell is the rectangle $0132$ pictured in Fig.~\ref{fig:scheme}~(\emph{Top}); the coordinates of the point in the flat state ($\phi=0$) are given in the following table:

\renewcommand{\arraystretch}{1.2}
\begin{center}
\begin{tabular}{cccc}
point & $x$ & $y$ & $z$ \\ \hline \hline
0 & $-1/2$ & 0 & 0 \\ \hline
1 & $1/2$  & 0 & 0 \\ \hline
2 & $-1/2$ & 0 & $\lambda$ \\ \hline
3 & $1/2$  & 0 & $\lambda$ \\ \hline
\end{tabular}
\end{center}

\begin{figure}[b]
\begin{center}
\includegraphics[width=.45\linewidth]{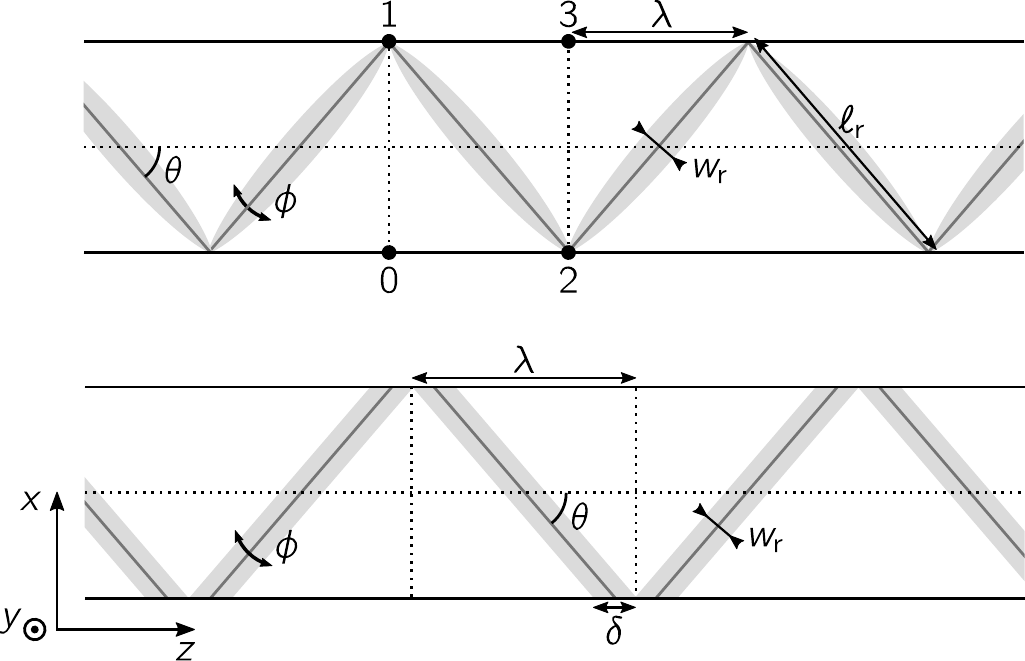}
\end{center}
\caption{Notations for the facetted twisted ribbon. The ridges are denoted by the gray areas. \emph{Top:} the stretching ridges do not affect the gross morphology. \emph{Bottom:} with cylindrical (isometric) ridges, a ``gap'' $\delta=w\ind{r}/\sin(\theta)$ is introduced between neighboring facets and the twist-contraction relation is modified.}
\label{fig:scheme}
\end{figure}

We first compute the transformation of the unit cell. We perform a rotation of angle $\phi$ of the point 3 around the edge 12, the points $012$ being fixed.

The rotation of a vector $\vv$ of an angle $\phi$ around an axis given by the unit vector $\hat \pp$ is given by:
\begin{equation}
\vv'=\mathcal{R}_{\hat\pp,\phi}=\vv_\parallel + \cos(\phi)\vv_\perp + \sin(\phi)\hat\pp\times\vv_\perp,
\end{equation}
where the parallel and perpendicular components are given by
\begin{align}
\vv_\parallel & = (\vv\cdot\hat\pp)\hat\pp,\\
\vv_\perp & = \vv - \vv_\parallel.
\end{align}

Here, the unit vector is
\begin{equation}
\hat\pp=\begin{pmatrix} - \sin(\theta) \\  0 \\ \cos(\theta) \end{pmatrix}.
\end{equation}
The vector to be rotated is
\begin{equation}
\vv_{23}=\begin{pmatrix}
1 \\ 0 \\ 0
\end{pmatrix},
\end{equation}
so that
\begin{align}
\vv_{23,\parallel} &= \sin(\theta) \begin{pmatrix} \sin(\theta) \\ 0 \\ - \cos(\theta) \end{pmatrix},\\
\vv_{23,\perp} & = \cos(\theta) \begin{pmatrix} \cos(\theta) \\ 0 \\ \sin(\theta) \end{pmatrix},\\
\hat\pp\times \vv_{23,\perp} & = \cos(\theta) \begin{pmatrix} 0 \\ 1 \\ 0 \end{pmatrix}.
\end{align}
Finally,
\begin{equation}
\vv_{23}'=\mathcal{R}_{\hat\pp,\phi} \vv_{23}= \begin{pmatrix}
1-\cos(\theta)^2[1-\cos(\phi)] \\ \cos(\theta)\sin(\phi) \\ -\cos(\theta)\sin(\theta)[1-\cos(\phi)]
\end{pmatrix}.
\end{equation}

The edges $01$ and $23$ are parallel in the flat state, but they are not in the twisted configuration. However, they are both orthogonal to the axis of the ribbon, which is no longer given by $\hat\ee_z$. Thus, they can be used to determine the axis $\aa$ of the ribbon: 
\begin{equation}
\aa=\vv_{01}\times\vv_{23}' = \begin{pmatrix} 1 \\ 0 \\ 0 \end{pmatrix}\times \begin{pmatrix}
1-\cos(\theta)^2[1-\cos(\phi)] \\ \cos(\theta)\sin(\phi) \\ -\cos(\theta)\sin(\theta)[1-\cos(\phi)]
\end{pmatrix} =  \cos(\theta) \begin{pmatrix}
0 \\ \sin(\theta)[1-\cos(\phi)] \\\sin(\phi) \end{pmatrix}.
\end{equation}
The magnitude of the axis vector $\aa$ is related to the angle $\alpha$ between the edges $01$ and $23'$: $|\aa| = |\sin(\alpha)|$. This angle is related to the twist per unit length $\eta$ through the length of a facet, which is $\lambda=1/\tan(\theta)$. Finally
\begin{equation}
|\aa|=\sin \left(\frac{\eta}{\tan(\theta)} \right).
\end{equation}
With the magnitude of $|\aa|$, this gives
\begin{equation}\label{eq:facets_twist}
\eta = \tan(\theta)\arcsin \left(\cos(\theta)\sqrt{\sin(\theta)^2[1-\cos(\phi)]^2+\sin(\phi)^2} \right).
\end{equation}
This expression differs from the one given in~\cite{bohr_13}, $\eta=\phi\sin(\theta)$ (Eq.~(22) with $\Upsilon=0$, $2\pi\Phi D/L=\eta$). 
The twist $\eta$ used here is defined as the angle $\Theta$ between the two clamps divided by the length $L$ of the ribbon. This quantity does not depend on the configuration of the ribbon: it is a topological invariant, the linking number~\cite{berger_06}.
On the other hand, Eq.~(22) in \cite{bohr_13} is for the total twist.

The contraction is obtained by looking at the distance between the facets along the axis of the ribbon:
\begin{equation}\label{eq:facets_contraction}
1-\chi = \hat\aa\cdot \hat\vv_{02}=\frac{\sin(\phi)}{\sqrt{\sin(\theta)^2[1-\cos(\phi)]^2+\sin(\phi)^2}}.
\end{equation}
It is in agreement with Eq.~(21) of~\cite{bohr_13}.

\subsubsection{Limits}

In the limit of small twist, the angle $\phi$ between neighboring facets is small: $\phi\to 0$. Taylor expansions of the twist (Eq.~(\ref{eq:facets_twist})) and the contraction (Eq.~(\ref{eq:facets_contraction})) give
\begin{align}
\eta & = \sin(\theta) \phi-\frac{\sin(\theta)^3}{24}\phi^3 - \left[  { ( 7 + 9 \cos(2 \theta)) \sin(\theta)^3 \over 3840} \right] \phi^5 + \mathcal{O}(\phi^7), \label{eq:taylor_twist}\\
\chi & = \frac{\sin(\theta)^2}{8}\phi^2 + \left[\frac{\sin(\theta)^2}{48}-\frac{3\sin(\theta)^4}{128} \right]\phi^4 + \mathcal{O}(\phi^6).\label{eq:taylor_contraction}
\end{align}
Inverting Eq.~(\ref{eq:taylor_twist}) and using it in Eq.~(\ref{eq:taylor_contraction}) leads to
\begin{equation}\label{eq:twist_contraction_smalltwist}
\chi = \frac{\eta^2}{8} + \left[\frac{1}{48\sin(\theta)^2}-\frac{5}{384} \right]\eta^4 + \mathcal{O}(\eta^6).
\end{equation}
At order $\eta^2$, the contraction does not depend on the size of the facets, which is given by $\theta$, and it is the same as the contraction of the wrinkled helicoid (main text, Eq.~(11)).
At order $\eta^4$, the contraction is an increasing function of the size of the facets, and the contraction of the helicoid is recovered in the limit of very small facets.

In the limit of small facets, $\theta\to \pi/2$, the twist and contraction are given by
\begin{align}
\eta &\underset{\theta\to\pi/2}{\longrightarrow} \sqrt{[1-\cos(\phi)]^2+\sin(\phi)^2}=2\sin(\phi/2),\\
1-\chi &\underset{\theta\to\pi/2}{\longrightarrow} \frac{\sin(\phi)}{\sqrt{[1-\cos(\phi)]^2+\sin(\phi)^2}}=\cos(\phi/2),
\end{align}
and the limiting values satisfy
\begin{equation}
\chi = 1-\sqrt{1-\frac{\eta^2}{4}},
\end{equation}
which is the contraction of the helicoid.
Rather naturally, the helicoid is recovered in the limit of small facets.

\section{Facets separated by isometric ridges (FIR)}

While minimal ridges are singular at the edges and do not alter the morphology of perfect facets (\emph{i.e.}, with sharp ridges), cylindrical ridges introduce a gap $\delta$ between neighboring facets (see Fig.~\ref{fig:scheme}). 
If $R\ind{c}$ is the radius of curvature of the ridges, their width is $w\ind{r}=\phi R\ind{c}$ and $\delta = w\ind{r}/\sin(\theta)=\phi R\ind{c}/\sin(\theta)$.
Here, we evaluate how these ridges modify the twist-contraction relation, Eq.~(\ref{eq:twist_contraction_smalltwist}), in the limit of small radius of curvature $R\ind{c}$.

The ridges modify both Eqs.~(\ref{eq:taylor_twist}) and (\ref{eq:taylor_contraction}). 
We use the result of Bohr and Markvorsen for the contraction (Eq.~(21) in Ref.~\cite{bohr_13}, which we expand in powers of $\phi$ and of the radius of curvature, leading to
\begin{equation}\label{eq:cyl_chi}
\chi = \frac{\sin(\theta)^2}{8}\phi^2 -\frac{\sin(\theta)^2\phi^3}{12\cos(\theta)}R\ind{c} + \left[\frac{\sin(\theta)^2}{48}-\frac{3\sin(\theta)^4}{128} \right]\phi^4 - \left[{R_c \over 960} {\left( 3 + 5 \cos(2 \theta) \right)  \sin(\theta)^2 \over \cos(\theta)}\right] \phi^5 + {\mathcal{O}(\phi^6)}. 
\end{equation}
To evaluate the effect of the twist, we use a simple geometrical argument: the fraction of the length that is in the ridges, $\delta/\lambda$, does not participate in the twist. 
Denoting $\eta_0(\theta,\phi)$ the twist for sharp ridges, the twist with finite size ridges is $\eta(\theta,\phi,R\ind{c})=(1-\delta/\lambda)\eta_0(\theta,\phi)$; using Eq.~(\ref{eq:taylor_twist}) for $\eta_0$ leads to 
\begin{align}
\eta(\theta,\phi,R\ind{c}) &= \left[1- \frac{\phi R\ind{c}}{\cos(\theta)}  \right] \left[\sin(\theta) \phi-\frac{\sin(\theta)^3}{24}\phi^3 - \left[  { ( 7 + 9 \cos(2 \theta)) \sin(\theta)^3 \over 3840} \right] \phi^5  + \mathcal{O}(\phi^7)\right], \\
& = \sin(\theta)\phi -\tan(\theta)R\ind{c}\phi^2 -\frac{\sin(\theta)^3}{24}\phi^3  + {R_c \over 24} \sin(\theta)^2 \tan(\theta) \phi^4 - \left[  { ( 7 + 9 \cos(2 \theta)) \sin(\theta)^3 \over 3840} \right] \phi^5 + \mathcal{O}(\phi^6 R\ind{c})
\end{align}
Inverting this relation to order $R\ind{c}$ and using it in Eq.~(\ref{eq:cyl_chi}) leads to
\begin{equation}
\chi=\frac{\eta^2}{8}+\frac{R\ind{c}\eta^3}{3\sin(2\theta)} + \left[\frac{1}{48\sin(\theta)^2}-\frac{5}{384} \right]\eta^4 + {R_c \over 240} \left( {13 + 5 \cos(2 \theta) \over \sin(\theta)^3 \cos(\theta)}\right) \eta^5 +\mathcal{O}(\eta^6).
\end{equation}
The change in the twist and the change in the contraction have the same order of magnitude, different signs, and finally the correction to the contraction as a function of the twist is positive.

%%We expect the wavelength to be much smaller than the width, so that the dominant contribution comes from bending in the longitudinal direction:
%\begin{equation}\label{eq:bending_scaling}
%U\ind{bend}\sim  B \left(\frac{A}{\lambda^2}\right)^2\sim   \frac{t^2 A^2}{\lambda^4}.
%\end{equation}
%Wrinkles relax the longitudinal compression, so that there is negligible strain in the longitudinal direction. 
%It is shown in the next subsection that the transverse displacement can be chosen to cancel the shear strain that could arise from non-uniform longitudinal displacement, so that the remaining stretching energy comes from stretching in the transverse direction.
%A displacement $A$ in the out-of-plane direction generates a strain scaling as $(A/W)^2=A^2$ in the transverse direction, leading to the stretching energy
%\begin{equation}
%U\ind{stretch}\sim Y \left(\frac{A}{W} \right)^4=A^4.
%\end{equation}

%Balancing the bending and stretching energies under the slaving condition Eq.~\ref{eq:slaving_scaling} leads to the following scalings for the wrinkles wavelength, amplitude and energy:
%\begin{align}
%\lambda & \sim (t/\eta)^{1/3}, \\
%A & \sim t^{1/3}\eta^{2/3},\label{eq:scaling_amplitude}\\
%U & \sim t^{4/3}\eta^{8/3}.
%\end{align}

%\bibliographystyle{unsrt}
%\bibliography{Facets_twisted_ribbons_PRL}{}

\begin{thebibliography}{10}
\bibitem{note_Rc}
Upon increasing tension, the ridges pinch at the ribbon's edges that generate stretching ridges. The assumption $R_{\mathrm{c}} \ll \eta$ that allows to discard the $\eta^3 R_{\mathrm{c}}$ term in $\chi$, clearly holds here. As shown in~\cite{witten_95}, the radius of curvature at the pinch $R_{\mathrm{c}} \le t$ and $\eta \gg t$ in the longitudinal buckled regime.
\bibitem{contraction}
For the helicoid shape, the contraction is given by $\chi = \eta^2/24 -T$~\cite{chopin_15}. Imposing a fixed length, i.e., $\chi =0$ requires $T = \eta^2/24$.
\end{thebibliography}

%\begin{thebibliography}{10}

%\bibitem{bohr_13}
%J. Bohr and S. Markvorsen.
%\newblock {{PLoS ONE}} {\bf 8}, 74932 (2013).

%\end{thebibliography}

%\end{document}

%

\end{widetext}

\end{document}